\documentclass[twocolumn,preprintnumbers,amsmath,amssymb]{revtex4-1}
\usepackage{epsfig}
\usepackage{color}
\usepackage{amsmath}
\begin{document}
\title{A simulation study of homogeneous ice nucleation in supercooled salty water}

\author{Guiomar D. Soria$^{\dagger}$, Jorge R. Espinosa$^{\dagger}$, Jorge Ramirez$^{\mathsection}$,Chantal Valeriani$^{\dagger,\ddagger}$, Carlos Vega$^{\dagger}$ and Eduardo Sanz$^{\dagger}$}
\affiliation{
$^{\dagger}$Departamento de Quimica Fisica I,
Facultad de Ciencias Quimicas, Universidad Complutense de Madrid,
28040 Madrid, Spain.\\
$^{\ddagger}$Departamento de Fisica Aplicada I,
Facultad de Ciencias Fisicas, Universidad Complutense de Madrid,
28040 Madrid, Spain.\\
$^{\mathsection}$ Departamento de Ingenieria Quimica Industrial y Medio Ambiente, Escuela Tecnica Superior de Ingenieros Industriales, Universidad Politecnica de Madrid, 28006 Madrid, Spain\\}

\date{\today}

\begin{abstract}

We use computer simulations to investigate the effect of salt on homogeneous
ice nucleation.
The melting point of the employed solution model was
obtained both by direct coexistence simulations and by thermodynamic
integration from previous calculations of the water chemical potential.
  Using a Seeding approach, in which we simulate ice seeds
embedded in a supercooled aqueous solution, we compute the nucleation rate
as a function of temperature for a 1.85  NaCl mole per water kilogram solution at 1 bar.  To improve
the accuracy and reliability of our calculations we combine Seeding with the
direct computation of the ice-solution interfacial free energy at coexistence
using the Mold Integration method.    We
compare the results with previous simulation work on pure water to understand
the effect caused by the solute.  The model captures the experimental trend that the
nucleation rate  at a given supercooling decreases when adding salt.  Despite
the fact that the thermodynamic driving force for ice nucleation is higher 
for salty water for a given supercooling, the nucleation rate slows
down with salt due to a significant increase of the ice-fluid interfacial free
energy.  The salty water model predicts an ice nucleation rate that is in good
agreement with experimental measurements, bringing confidence in the predictive
ability of the model.

\end{abstract}
\maketitle

\section{Introduction}

The formation of ice from supercooled water is arguably the most important
freezing transition on Earth.  Such transition has important consequences in
geology \cite{weathering_book}, food industry \cite{bacterial_ice_nucleation,maki74}, or cryopreservation \cite{cryopres}. 
Most water on Earth contains dissolved salt. Therefore, understanding freezing
in salty water is of out-most importance. 
In this work we focus on homogeneous nucleation, the case in which ice starts growing in the 
bulk supercooled liquid. 
Salt hinders ice formation both by
decreasing its melting temperature, $T_m$, and by increasing the
supercooling with respect to $T_m$ required to observe freezing \cite{kannoJPC1977}.  
We aim at understanding the latter effect in this paper. 

The first step of the freezing transition 
is the nucleation 
of a critical crystal cluster --one that has equal chances
to melt or irreversibly grow \cite{kelton}.
The number of such critical clusters appearing per unit time and volume is the nucleation 
rate, $J$.
There are several
experimental works where $J$ has been measured for ice in salty water \cite{kannoJPC1977,alpertNaClaqueous}. 
While the homogeneous nucleation rate can be experimentally measured in careful experiments 
to avoid the presence of impurities, 
other relevant nucleation parameters such as the size or the structure of the
cluster, or the ice-liquid interfacial free energy are difficult to obtain in experiments \cite{ickespccp2015}.

Molecular simulations are an excellent complement to experimental studies of
the freezing transition because they have access to detailed information at the
molecular scale \cite{reviewMichaelides2016}.  There are many simulation 
studies of ice nucleation in pure
water (e.g. \cite{JACS_2003_125_07743,quigley:154518,geigerJCP2013,valeria_pccp_2011,Malkinpnas2012,galli_mw,haji-akbariPNAS2015,reviewMichaelides2016,jacs2013}), 
and salt precipitation in supersaturated solutions \cite{chakrabortyJPCL2013,lanaroNaCl2016,PRL_2004_92_040801,petersJACS2015}, but not much work has been devoted to study freezing of 
salty water \cite{condePCCP2017,bullockFD2013,icenucNaClJung2008}.  In the few existing studies, 
neither nucleation rates nor ice-solution interfacial free energies
have been calculated. 

Like salt, pressure is known to slow down ice nucleation \cite{kannoScience1975,kannoJPC1977}.
After having studied the effect of applying high
pressure on homogeneous ice nucleation \cite{espinosaPRL2016}, we
performed a comparative study of the effects of pressure and salt where we
concluded that both factors hinder ice nucleation by increasing the ice-liquid
interfacial free energy \cite{espinosaJPCL2017}. In this paper we focus the
discussion on the effect of salt alone and provide details of our study of
homogeneous ice nucleation in salty solutions that were not given in Ref.
\cite{espinosaJPCL2017}. 

We study ice nucleation in a 1.85 NaCl mole per water kilogram solution (1.85 m) aqueous solution using 
a model that combines  TIP4P/2005
water \cite{JCP_2005_123_234505} and the Joung Cheetham/SPC/E NaCl \cite{JCNaClpotential} 
model \cite{analaura}. 
 
We attempt to rationalise the nucleation rate by examining
the variables upon which it depends according to Classical Nucleation Theory \cite{kelton,ZPC_1926_119_277_nolotengo,becker-doring}. 
First, we determine the melting temperature of ice in a 1.85 m NaCl solution at
normal pressure via two different routes, one based on direct coexistence
simulations between both phases \cite{laddcpl1977} and another based on
calculations of the chemical potential of the bulk phases separately
\cite{JCP_2007_126_014507,analaura}.  Using thermodynamic integration from the
melting point \cite{frenkelbook} we compute the chemical potential difference
between ice and water in solution.  We then obtain via seeded simulations the
size of critical ice clusters for three different temperatures
\cite{seedingvienes}.  By simulating a critical cluster we estimate the
frequency with which molecules attach to it \cite{Nature_2001_409_1020}.  To
improve the reliability of our calculations we also compute the interfacial
free energy at coexistence (a thermodynamic parameter that cannot
be reliably measured experimentally) with the Mold Integration method
\cite{espinosaJCP2014_2,pocillosagua}.  Combining the simulation results with
Classical Nucleation Theory (the seeding method \cite{seedingvienes}) 
we estimate the nucleation rate and the interfacial
free energy for a wide supercooling range. 
To rationalise the effect of salt on ice nucleation 
we compare these results with those previously 
obtained for homogeneous ice nucleation in pure water 
\cite{jacs2013,espinosaJCP2014}.

\section{Model}

We use the TIP4P/2005 model for water \cite{JCP_2005_123_234505}. For NaCl we use the Joung Cheetham model parametrised for SPC/E water
\cite{JCNaClpotential}, although in this work we combine
it with TIP4P/2005 water as in Ref. \cite{analaura}. 
Notice that it would not be a good idea to use SPC/E water as solvent for this
study because its melting point for pure water is about 215K (too
far from from the experimental value) and besides the kinetics for the
supercooling considered in this work (i.e up to 40K) would be terribly slow for
this model. 
The ion-water cross interaction follows the Lorentz-Berthelot combination rules \cite{lorentzrule,berthelotrule}
for the Lennard-Jones part and the Coulomb law for the electrostatic part. 
The solubility limit for this brine solution force field at 1 bar and 298 K is 3.5 m (NaCl mol per H$_2$O kg) \cite{analaura}, somewhat
smaller than the experimental value of 6.15 m \cite{solubNaCl}.

\section{Simulation details}

All runs are performed at constant pressure of $p = 1$~bar and NaCl concentration of 1.85 m.

All our simulations are run with the Molecular Dynamics (MD) GROMACS
package \cite{hess08}.   We use a
Parrinello-Rahman barostat\cite{parrinello81} with a relaxation
time of 0.5 ps to fix the pressure.  To keep the temperature constant we employ a velocity-rescale thermostat
\cite{bussi07} with a relaxation time of 0.5 ps.  The time step for the Verlet
integration of the equations of motion is 2 fs.  We use Particle Mesh Ewald
Summations \cite{CPL_2002_366_0537} to deal with electrostatic interactions.
The cut-off radius for both dispersive interactions and the real part of
electrostatic interactions is 9 \AA. The LINCS algorithm is used to 
fix the geometry of the water molecules \cite{lincs97,lincs2008}.

\section{Methods}

To obtain the melting temperature for the selected water-ion force field we use two different approaches. 
On the one hand we use 
the so-called direct coexistence method \cite{laddcpl1977}. Such method
consists in simulating the solid in contact with the liquid at several
temperatures \cite{bullockFD2013,condePCCP2017}. Below the melting temperature, the solid grows at the expense of
the fluid phase, and vice versa.  In this way the melting temperature can be
enclosed within a certain range.  The method has been used in the past by some
of us to obtain the melting temperature of pure water
\cite{garciafernandezJCP2005}, hard spheres \cite{noyaJCP2008,sigmoide} or  sodium
chloride \cite{aragones12}.
On the other hand, we perform thermodynamic integration \cite{frenkelbook} from 
previous calculations of the chemical potential of the solvent and the solute \cite{JCP_2007_126_014507,analaura} 
and from the melting point of pure water \cite{condeJCP2013} in order to find the melting 
temperature as that for which the chemical potential of water
in ice and in solution coincide.
The calculation of the chemical potential of the components of a solution 
is a difficult task that has recently received great attention in the context of the
computation of crystal solubilities \cite{mesterJCP2015,panagiotopoulosjcp2015,paluchJCP2010,mouckaJPCB2012,ivoreview2016,solubilityfrenkel2017}. 
 
Ice nucleation at moderate
supercooling is beyond the time scale of standard Molecular Dynamics 
simulations. Special techniques are needed to promote the nucleation (rare) event. 
In this work we use the 
seeding method, which has recently been carefully validated by us \cite{seedingvienes}.  
The seeding technique consists in inserting an ice cluster in the supercooled
fluid (salty water) and then simulating the resulting equilibrated configuration
at several temperatures
to obtain the  temperature at which the inserted cluster reaches a critical size. 
Such information, combined with Classical Nucleation Theory \cite{becker-doring,ZPC_1926_119_277_nolotengo,kelton} 
provides estimates of important quantities for the nucleation process such as the 
ice-liquid interfacial free energy \cite{carignano,baiJCP2006}, the height of the nucleation free energy
barrier \cite{hurtadoseedingising2008,Jacobson_molinero} and the nucleation rate \cite{knottJACS2012,jacs2013,espinosaJCP2014,espinosaJCP2015,zaragozaJCP2015,espinosaPRL2016}. 
The latter requires launching MD trajectories from the critical cluster in order to get the kinetic prefactor.

To compute the ice-solution interfacial free energy at coexistence (for a flat interface) we use
the Mold Integration method, that has been developed by some of us \cite{espinosaJCP2014_2} and employed 
to study the crystal-fluid interface for NaCl and water \cite{espinosaJCP2015,pocillosagua,espinosaPRL2016}.  
The method consists in gradually inducing the formation of a crystal slab in 
the fluid at coexistence conditions with the aid of a mold of potential energy 
wells placed at the lattice sites of a crystal plane \cite{espinosaJCP2014_2}. 
The work needed to form such slab can be obtained by thermodynamic integration 
and is directly related to the crystal-fluid interfacial free energy \cite{espinosaJCP2014_2}. 

We refer the reader to Refs. \cite{noyaJCP2008} (direct coexistence), \cite{seedingvienes} (seeding)
and \cite{espinosaJCP2014_2} (Mold Integration) for a detailed description of the employed
methods.  Nevertheless, in the following
section we briefly explain these methods as we present and discuss
the results. 

\section{RESULTS}
\subsection{Melting temperature}

\begin{figure*}[!htb]
\begin{center}
\includegraphics[clip,width=0.6\textwidth]{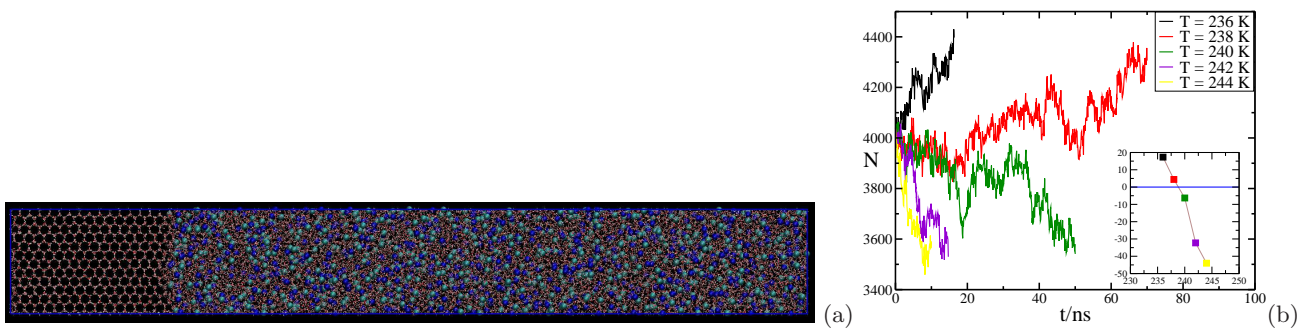} (a)
\includegraphics[clip,width=0.3\textwidth]{./coex_dir.eps}(b)
\caption{(a) Snapshot of the starting configuration used to evaluate the coexistence temperature
between ice Ih and the NaCl aqueous solution. The solution is elongated in order to minimize
concentration changes as the ice slab grows/shrinks.(b) Main: number of molecules belonging to the 
ice slab as a function of time at different temperatures. Inset: slope of a linear fit to the curves in the main panel as a function
of temperature. The melting temperature is taken as that for which the interpolated slope is zero (239K).}
\label{DC}
\end{center}
\end{figure*}

To start with, we compute the melting temperature, $T_m$
of ice in contact with the NaCl solution at the chosen 
NaCl concentration (1.85 m). We perform such calculation 
with two different methods: direct coexistence and 
and thermodynamic integration.

\subsubsection{Direct coexistence}
In the direct coexistence method an ice and a 1.85 m NaCl solution slab are put 
at contact as shown in Fig. \ref{DC} (a).
Simulations are carried out in the anisotropic isothermic-isobaric NpT ensemble \cite{sigmoide}. 
When the imposed temperature is below $T_m$ 
the ice slab grows, and vice versa.  In order to know whether the ice slab
grows or melts we monitor $N$, the number of molecules in the ice phase 
(see appendix \ref{orderparam}). In Fig. \ref{DC}(b) we plot $N$ versus time for 
different temperatures. At 238 K the number of ice-like particles grows whereas
it decreases at 240 K: $T_m$ is thus $239\pm1 K$. 

In order to use an unambiguous criterion to determine $T_m$ we compute the
slope of a linear fit to each $N(t)$ trajectory in Fig. \ref{DC}(b). If the slope is positive, ice 
is growing and viceversa. Therefore, the temperature at which the slope
is zero   corresponds to $T_m$. The slope (in molecule/ns units) as a function
of temperature is shown in the inset 
of  Fig. \ref{DC}(b). As announced, the slope becomes 0 at $T_m = 239 K$. 

When estimating $T_m$ with the direct coexistence method, 
the salt concentration may change as water molecules are
added/removed to/from the solution due the ice melt/growth. This is an undesirable 
finite size effect because we are interested in obtaining $T_m$ for a specific value
of the concentration (1.85 m). To alleviate such finite size effect we took three measures:
(i) use a large system size (23100 water molecules, of which 4050 were in the ice phase + 638 NaCl) (ii) use a solution 
slab much larger than the ice slab (see Fig. \ref{DC} (a)) and (iii) determine
whether ice grows or melts when only a few hundred water molecules melt or freeze. 
Proceeding with such caution we estimate that the NaCl concentration in solution never 
changes more than 1\% from its original value.

\subsubsection{Thermodynamic integration}

Another route to obtain the melting temperature is
by finding the point at which the ice chemical potential ($\mu_w^i$) 
equals the chemical potential of water
in solution ($\mu_w^{sol}$). 
To find such point we compute separately $\mu_w^i-\mu_w^0$ and
$\mu_w^{sol}-\mu_w^0$, where $\mu_w^0$ is the chemical potential
of pure water. The temperature at which these two chemical potential
differences become equal is the melting point.  
To obtain chemical potential differences we use the
Gibbs-Helmholtz equation:
\begin{equation}
\left(\frac{\partial(\mu^{\alpha}_{w}/T)}{\partial T}\right)_p=-\frac{\bar{h}^{\alpha}_{w}}{T^2},
\label{gh}
\end{equation} 
Where the superscript $\alpha$ stands for a given phase (pure water, $w$, ice, $i$, or solution, $sol$). 

\begin{figure}[!htb]
\begin{center}
	\includegraphics[clip,width=0.4\textwidth]{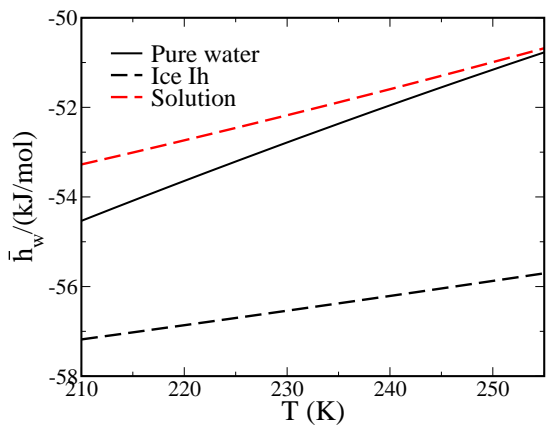}
\caption{Molar enthalpies 
of ice, pure water, and partial molar enthalpy of water in the studied NaCl solution
as a function of temperature.}
\label{enthalpies}
\end{center}
\end{figure}

$\bar{h}^{\alpha}_{w}$ is the water partial molar enthalpy which, for ice or pure water, is simply the molar enthalpy.
For the solution, however, $\bar{h}^{sol}_{w}$ is defined as:
\begin{equation}
	\bar{h}^{sol}_{w}=N_A\left(\frac{\partial H}{\partial N_{w}}\right)_{N_{NaCl,p,T}}\approx N_A\left(\frac{\Delta H}{\Delta N_{w}}\right)_{N_{NaCl,p,T}}
\end{equation}
where $N_A$ is the Avogadro's number and $H$ is the system's enthalpy. 
We numerically evaluate the derivative above by computing the enthalpy for two different
systems that have the same number of NaCl ion pairs and a different number of water molecules. 
It is important to make sure that the NaCl concentration in both systems (1.836 and 1.863 m in our case) 
narrowly encloses the concentration of interest (1.85 m). 
By dividing the enthalpy difference, $\Delta H$, by the difference of the number of 
water molecules, $\Delta N_{w}$, we get an estimate of $\bar{h}^{sol}_{w}$ at 1.85 m.
Obtaining $\Delta H$ requires running long simulations (from 800 ns at 190 K to 100 ns at 300 K) because the enthalpy difference 
between both systems is very small.  
In Fig. \ref{enthalpies} we plot $h_{w}$  as a function of temperature for the solution (red 
dashed line) and the ice phases (black dashed line) and compare it
with that of pure water (black solid line). 
As expected, the enthalpy of water in the ice phase is the lowest, whereas 
the enthalpy of pure water is lower than the partial molar enthalpy of water in the solution, reflecting the 
fact that the hydrogen bond network is disrupted by the ions.  

Once the
temperature dependence for the (partial) molar enthalpy is known, 
the water chemical potential difference between phases
$\alpha$ and $\beta$ at temperature $T_B$ can be obtained by
integrating  
Eq. \ref{gh} from $T_A$, a temperature of known chemical 
potential difference, to $T_B$:
\begin{equation}
\frac{\mu_{w}^{\alpha}(T_B)-\mu_{w}^{\beta}(T_B)}{T_B}=
\frac{\mu_{w}^{\alpha}(T_A)-\mu_{w}^{\beta}(T_A)}{T_A}-\int_{T_A}^{T_B}\frac{\bar{h}^{\alpha}_{w}-\bar{h}^{\beta}_{w}}{T^2}dT
\label{intgh}
\end{equation}

\begin{figure}[!htb]
\begin{center}
\includegraphics[clip,width=0.4\textwidth]{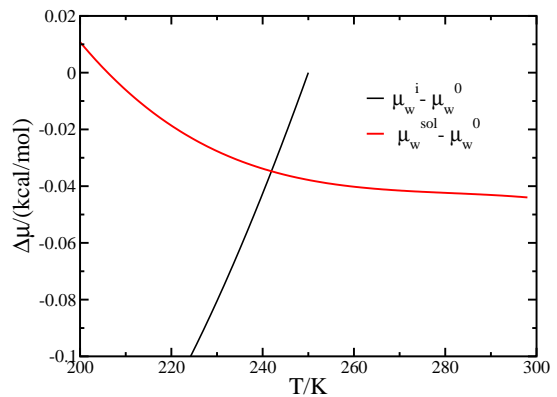}
\caption{Chemical potential differences (see legend) as a function 
of temperature. The crossing point corresponds to the melting 
temperature.}
\label{tmint}
\end{center}
\end{figure}

For $\alpha=i$ and $\beta=0$ (corresponding to ice and pure water) we use $T_A = T_m^0$, 
the melting temperature of pure water (250 K for TIP4P/2005), 
at which the chemical potential difference is 0. 
$\mu_w^{i}-\mu_w^{0}$ as a function of temperature
is shown in Fig. \ref{tmint} (black line) 
and has been previously reported by some of the authors \cite{espinosaJCP2014,jacs2013}.

For $\alpha=sol$ and $\beta=0$ we use $T_A=298 K$, where the
chemical potential difference $\mu_w^{sol}-\mu_w^{0}$ is equal
to $RT\ln a_w(298K)$. The activity of water at 298 K, $a_w(298K)$, 
has been reported in Ref. \cite{analaura} as a function of the concentration
for the same solution model used here. 
The value we take for 1.85 m is 0.93.
The red curve in Fig. \ref{tmint}
is $\mu_w^{sol}-\mu_w^{0}$.

The point where both curves in Fig. \ref{tmint} cross corresponds to the melting temperature, 242
K. This value is different, but within the error bar, of that obtained by means
of direct coexistence, 239 K.  The error in the direct coexistence method comes
from the stochasticity associated with the finite size of the system
\cite{sigmoide}.  We have used the average value between direct coexistence and
the chemical potential route, 240.5 K, as the melting temperature for all
calculations in the paper.

\subsection{Critical cluster size, $Nc$}


\begin{figure}[!htb]
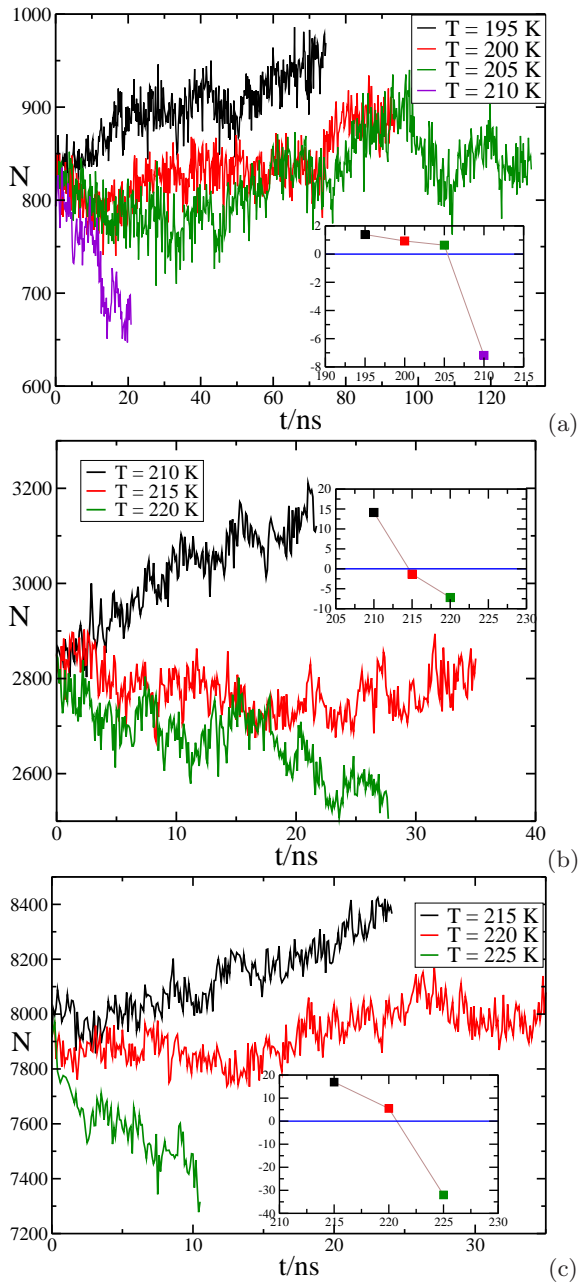

\begin{center}
\includegraphics[clip,width=0.4\textwidth]{./big.eps}(a)
\includegraphics[clip,width=0.4\textwidth]{./large.eps}(b)
\includegraphics[clip,width=0.4\textwidth]{./huge.eps}(c)
\caption{Time evolution of the number of particles in the ice cluster for different inserted
clusters and temperatures. Inset: Slope of a linear fit of N(t) as a function of temperature. 
The temperature at which the inserted cluster is critical is taken as that at which the 
interpolated slope is zero. The size of the inserted clusters 
and the temperatures at which they are found to be critical are reported in table \ref{tabla}.}
\label{nvst}
\end{center}
\end{figure}

The size of the critical clusters is determined simulating ice seeds embedded in a 
supercooled solution. All inserted seeds have spherical shape and ice-Ih structure. 
Spherical
ice Ih clusters (or staking mixtures of ice-Ih and ice-Ic) seems to be
the nucleation pathway followed 
in homogeneous ice nucleation from pure water \cite{jacs2013,zaragozaJCP2015,espinosaPRL2016}. 
We insert pure ice seeds with no ions in the crystal lattice. 
This approximation is
inspired by the fact that the brine coexists with pure ice in the experimental phase diagram \cite{koopNaClphasediag}.
To check this approximation we have
performed a 260 ns direct coexistence simulation below
the melting temperature and computed the fraction of ions coming
into the newly grown ice lattice.  Such fraction was smaller than 0.2 per cent. 
This value, which is consistent with previous simulation 
work \cite{condePCCP2017,icenucNaClJung2008}, is sufficiently small
to justify the approximation of not including ions in the ice seeds.

In Ref. \cite{espinosaJCP2015} (Fig. 7)
we describe in some detail how to equilibrate an initial configuration
of an embedded crystal seed into a supercooled fluid to start up the calculation. In Table \ref{tabla} we give details on the
three different initial configurations prepared ranging from a cluster of about 800 molecules up to one about 10 times larger. Starting from 
an equilibrated configuration
with an ice cluster of $N_c$ water molecules we launch several isotropic $NpT$
simulations and monitor the number of ice molecules in the cluster, $N$. 
Such number is determined as explained in appendix \ref{orderparam}.
If $N$ decreases, the inserted cluster is
subcritical at the chosen temperature, whereas if $N$ increases the cluster
is post-critical. In such way we can determine the temperature at which the
inserted cluster is critical. This is similar in spirit to the way the melting
temperature is determined in direct coexistence simulations.  Again, the large
system sizes employed ensure that there are negligible concentration changes as
the cluster grows or melts, which is the experimentally relevant case for ice
nucleation in salty solutions at constant pressure and temperature. 

\begin{figure*}[!htb]
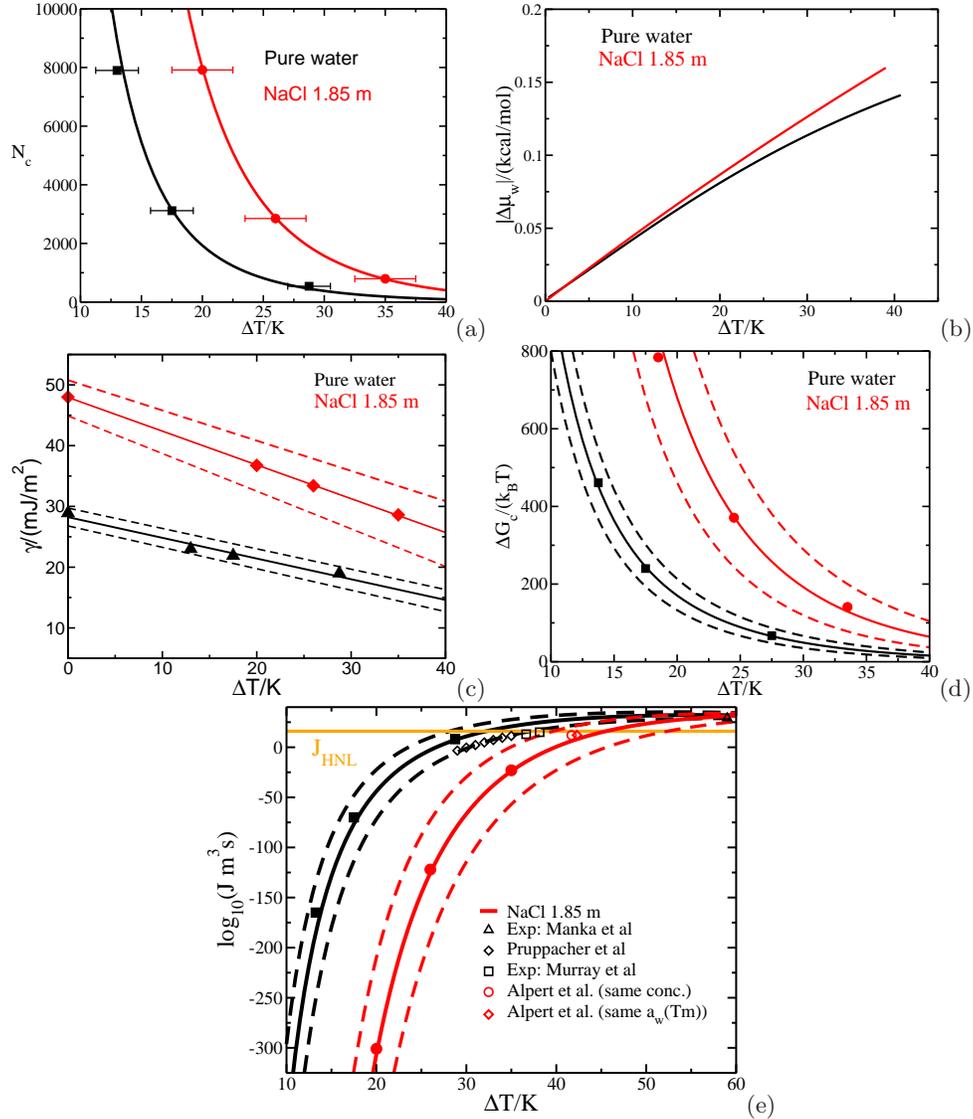

\begin{center}
\includegraphics[clip,width=0.33\textwidth]{./Nvssupercooling_forlongpaper.eps}(a)
\includegraphics[clip,width=0.33\textwidth]{./Deltamu_forlongpaper.eps}(b)
\includegraphics[clip,width=0.33\textwidth]{./gamma_finales_seeding_mas_pocillos_forlongpaper.eps}(c)
\includegraphics[clip,width=0.33\textwidth]{./deltasg.eps}(d)
\includegraphics[clip,width=0.4\textwidth]{./Curva_nucleacion_final_bien_forlongpaper.eps}(e)
\caption{Plotted for pure (black) and 1.85 m salty water (red) as a function of the supercooling:(a) Number of particles
in the critical cluster; (b) Water chemical potential difference between the liquid and the solid phases; (c) ice-water
interfacial free energy;  
(d) height of the ice nucleation free energy barrier; (d) decimal logarithm of the nucleation rate for pure water and the 1.85 m NaCl solution. 
$J_{HNL}$ is the homogeneous nucleation line corresponding to a rate of J=$10^{16} m^{-3}s^{-1}$ measured in typical experiments \cite{koopNature2000}.
Solid symbols are our seeding results and empty symbols are experimental results by Pruppacher \cite{pruppacher1995}, Murray et al. \cite{murray2010}, Alpert et al. \cite{alpertNaClaqueous} 
and Manka et al. (triangle at the top right of the figure) \cite{manka2012}
as indicated in the legend.}
\label{various}
\end{center}
\end{figure*}

In Figs. \ref{nvst}(a), (b) and (c) we show $N$ versus time for the three simulated
ice clusters at different temperatures. As for the determination of $T_m$, 
from the temperature at which the slopes of linear fits to $N(t)$ interpolate to 0 we 
determine the temperature at which the inserted ice cluster is critical. 
The slope of $N(t)$ as a function
of temperature is shown in 
the insets of Figs. \ref{nvst}(a), (b) and (c). The results for the temperature at which 
the clusters are found to be critical are reported in Table \ref{tabla} 
and shown in Fig. \ref{various}(a) as a function of the supercooling, $\Delta T= T_m -T$.
For comparison, we also include in the figure the results for pure water refined from our previous work with more statistics 
in the seeding simulations \cite{jacs2013,espinosaJCP2014}.
As shown in Fig. \ref{various}(a),
for a given supercooling, the number of particles needed to form a critical cluster is larger
in solution than in pure water. This suggests that, apart from the trivial effect of lowering $T_m$,
the presence of salt hinders the nucleation of ice by changing the thermodynamic or the kinetic
parameters that affect ice nucleation.
 
\begin{figure}[!htb]
\begin{center}
\includegraphics[clip,width=0.4\textwidth]{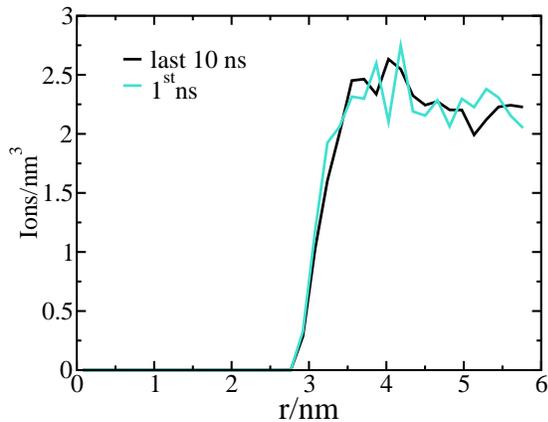}
\caption{Ion density as a function of the radial distance from the center of mass of the ice cluster. Turquoise, density profile averaged over
the first nanosecond of the trajectory in Fig. \ref{nvst}(b) corresponding to 215 K. Black, density profile averaged over the last 10 ns of the same trajectory.}
\label{idp}
\end{center}
\end{figure}

To make sure that the interface between the inserted 
ice cluster and the solution is properly equilibrated at the beginning 
of the simulations shown in Fig. 4 we examine the density 
profile of the ions along the radial distance from the ice 
cluster centre of mass.
We analyse the trajectory corresponding 
to 215 K in Fig. 4(b) because the cluster size stayed 
roughly constant at about 2800 molecules throughout the 
run. In Fig. \ref{idp} we compare the ions density profile in the 
first nanosecond of the trajectory (turquoise curve) to 
that corresponding to the last 10 ns (black curve). 
The ion density profile is 0 in the interior of the cluster and increases at
the interface up to the equilibrium bulk density: the ice cluster is
characterised by a 3 nm radius and an interface width of nearly 1 nm. 
The 
ion density is surely equilibrated in the later period, 
starting at 25 ns of the trajectory,
because the time required for ions to diffuse their own diameter 
at 215 K is about 2.5 ns. It is clear from Fig. \ref{idp} that both 
density profiles are quite similar, both at the interface and 
away from the ice cluster, which proves that we have 
started the trajectory from a configuration where the ice 
cluster is surrounded by an equilibrium distribution of ions.

\begin{table*}
\begin{center}
\begin{tabular}{ccccccccccc}
System & $N_{c}$ & $T_{c}$ (K) & $\Delta T$ (K) & $\Delta \mu$ (kcal/mol) & $\rho_w$/(g/cm$^3$) & $\gamma$ (mN/m) & $\Delta G_{c}/(k_{B}T)$ & $f^{+}/s^{-1}$ & $\log_{10}$(J($m^{-3}s^{-1}$)) \\ \hline
Sol     & 794     & 205.5   & 35.0       & 0.145    & 0.980 &     28.6     & 141            & $2.69x10^{10}$ & -23           \\
Sol     & 2850    & 214.5   & 26.0       & 0.111    & 0.982 &     33.4     & 371            & $2.10x10^{11}$ & -122          \\
Sol     & 7916    & 220.5   & 20.0       & 0.087    & 0.983 &     36.8     & 784            & $8.40x10^{11}$ & -301          \\
W      & 539     & 221.25   & 28.75      & 0.110    & 0.961 &     18.9     & 67             & $1.30x10^{11}$ &  8            \\
W      & 3117    & 232.5    & 17.5       & 0.071    & 0.974 &     21.9     & 240            & $1.22x10^{12}$ & -66           \\
W      & 7900    & 237.0    & 13.0       & 0.055    & 0.978 &     23.0     & 461            & $3.00x10^{12}$ & -163          \\
\end{tabular}
\caption{Variables involved in the calculation of the ice Ih nucleation rate in pure and salty water.
See main text for the meaning of all variables, whose values are reported with the corresponding units.
The melting temperatures at 1 bar for pure water and the studied 1.85 m NaCl solution are 250 and 240.5 K, respectively.}
\label{tabla}
\end{center}
\end{table*}

\subsection{Driving force for ice nucleation}

According to Classical Nucleation Theory, the free energy change associated to the formation of a crystal 
cluster with $N$ molecules is given by two competing terms:
\begin{equation}
	\Delta G (N)=-N|\Delta \mu_{w}|+\gamma A
	\label{cnt}
\end{equation}
The first term favours the formation of the cluster and takes into account the fact that, below $T_m$, 
the chemical potential of water in ice is lower than in solution. 
Therefore, the chemical potential difference of water in both phases, $|\Delta \mu_{w}|=|\mu_w^{i}-\mu_w^{sol}|$ is 
the thermodynamic driving force for ice nucleation. 
The second term in the equation above hinders the nucleation of the crystal and is the product of the area of the cluster's surface, $A$, and the
ice-solution interfacial free energy, $\gamma$. 

To obtain $|\Delta \mu_{w}|$ as a function 
of temperature we use Eq. \ref{intgh} 
with $\alpha = sol$, $\beta = i$ 
and $T_A=T_m=240.5$ K, where $\mu_w^{sol}-\mu_{w}^i =0$. 

The chemical potential difference is plotted in Fig. \ref{various}(b), where we compare with our previous 
results for pure water. For a given supercooling, the thermodynamic 
driving force for the formation of ice is larger in solution (red)
than in pure water (black), specially at high supercooling. 
This is consistent with the fact that the partial enthalpy of water in solution is higher than the enthalpy
in pure water (Fig. \ref{enthalpies}). Therefore, the free energy gain when a fluid water molecule becomes part
of the ice cluster is larger in solution than in pure water. According to $|\Delta \mu_{w}|$, then, 
the formation of ice in solution should be easier than in pure water, which is not consistent with Fig. \ref{various}(a),
where we show that, for a given supercooling, larger clusters are required to nucleate ice in solution than in pure water.
However, $|\Delta \mu_{w}|$ is not the full story. One also has to take into account the ice-solution interfacial 
free energy, $\gamma$, which is what we discuss in the following section.

\subsection{Ice-solution interfacial free energy}

\begin{figure}[!htb]
\begin{center}
        \includegraphics[clip,width=0.4\textwidth]{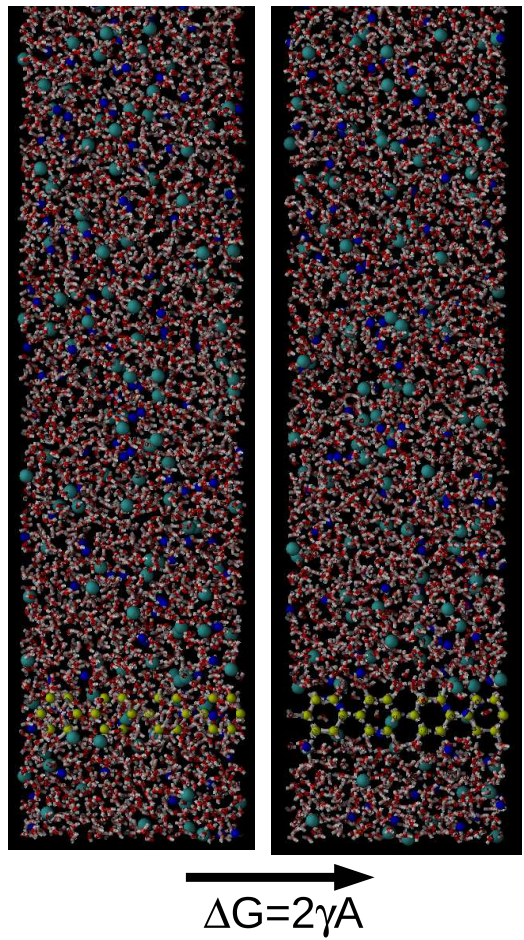}
\caption{Snapshot of a 1.85 m NaCl solution configuration at the melting temperature at 1 bar. The 
square well-like interaction between the mold sites (yellow particles) and the oxygens of the water molecules is
switched off in the left panel and on in the right panel. Note that the number for crystal planes in the mold, $N_l$, is three in this case.}
\label{fotopocillos}
\end{center}
\end{figure}

By maximizing Eq. \ref{cnt}  and assuming a spherical cluster shape the following expression can be found for the interfacial free energy:
\begin{equation}
\gamma = \left(\frac{3N_c\rho_s^2|\Delta \mu_{w}|^3}{32\pi}\right)^{1/3},
\label{gammaeq}
\end{equation}
which depends on the number of particles in the critical cluster, $N_c$, the ice number density, $\rho_s$, and
the ice-solution water chemical potential difference, $|\Delta \mu_{w}|$.
In Table \ref{tabla} we give the calculated values of $\gamma$ for the three studied cluster sizes. 
Both in the table and in Fig. \ref{various}(c) we show that the ice-solution interfacial free energy is significantly 
larger than the ice-water one (for a given $\Delta T$). 
Therefore, despite the fact that the bulk free energy drop associated to the formation of an
ice cluster is larger in solution than in pure water (Fig. \ref{various}(b)), the increase of surface free energy is higher for the solution (Fig. \ref{various}(c)), which
explains why salt hinders the nucleation of ice clusters (Fig. \ref{various}(a)).
In fact, once both $\gamma$ and $\Delta \mu$ are known, the height of the nucleation free energy barrier can be evaluated
according to CNT as:
\begin{equation}
\frac{\Delta G_c(T)}{k_BT} = \frac{16\pi(\gamma(T))^3}{3(\rho_s(T))^2|\Delta \mu_w(T)|^2k_BT}
\label{dgceq}
\end{equation}
The calculated values are reported in Table \ref{tabla} and plotted in Fig. \ref{various}(d). 
Clearly, for a given supercooling more work is required to reversibly grow an ice cluster in solution than in pure water. 
The responsible for such increase is the interfacial free energy, given that $\Delta \mu_w$ actually contributes to lower $\Delta G_c$. 

To make sure that $\gamma$ increases when adding salt we also compute $\gamma$
for a flat interface (at coexistence) using the Mold Integration method
\cite{espinosaJCP2014_2} that we have recently employed to compute the
ice-water interfacial free energy for pure water at ambient \cite{pocillosagua}
and high pressure \cite{espinosaPRL2016}.  In this method, which is only valid for
coexistence conditions, a mold composed of
square wells placed in the lattice sites of one or several crystal planes is gradually
switched on to induce the formation of a solid slab in the fluid at
coexistence conditions. The idea is sketched in Fig. \ref{fotopocillos}. We note that the interfacial
crystal halo generated by the mold is free to fluctuate and to incorporate and expel ions to reach
its equilibrium structure. The
free energy difference between the fluid and the fluid with the structure
induced by the mold can be computed by integrating the average number of filled
wells, $<N_{fw}>$, along such thermodynamic path: 
\begin{equation} 
\Delta G = N_w \epsilon_m - \int^{\epsilon_m}_{0} <N_{fw}> d\epsilon 
\label{eqpoc}
\end{equation} 
where
$N_w$ is the number of wells in the mold, $\epsilon$ is the well depth and
$\epsilon_m$ is the maximum well depth (10 $k_BT$ in our case).  When the well radious, $r_w$, is equal
to a certain ``optimal'' value, $r_w^o$, $\Delta G$ is equal to $2 \gamma A$
\cite{espinosaJCP2014_2}, where $A$ is the area of the mold (the factor of 2
comes from the fact that two interfaces are generated).  The position of the
wells is fixed during the simulation and only the side of the simulation box
perpendicular to the interface fluctuates to keep the pressure constant. 
Mold Integration can be easily implemented in GROMACS using a tabulated interaction
potential between the mold and the fluid particles \cite{espinosaJCP2014_2}.
A cut-off of 14 \AA~ is used for all interactions. 
 The
interfacial free energy of different crystal orientations can be obtained by
using molds corresponding to different crystal planes
\cite{espinosaJCP2014_2,pocillosagua}.  In this work we study the basal,
primary prismatic (pI) and secondary prismatic (pII) orientations.  
In table  \ref{tablapocillos} we report details on our
Mold Integration calculations such as $N_w$, $A$, $r_w^o$ or the number
of crystal planes in the mold, $N_l$, for each crystal orientation studied. 

\begin{table}[t]
\centering
\begin{tabular}{ccccccc}
\hline
System & Plane & $(L_y L_z)/\AA^2$ & $N_w$ &  $N_l$ & $r_w^0/$\AA & $\gamma/(mJ/m^2)$  \\
\hline
\hline
Sol & basal & 1127.7 & 192 & 3 & 0.97(10)  & 43(2) \\
Sol & pI & 1063.2 & 192 & 3 & 0.72(10) & 49(3)  \\
Sol & pII & 920.73 & 192 & 3 & 0.85(10) & 52(3) \\
Sol & $\gamma_0$ &  &  &  & & 48(3)    \\
\hline
W & basal & 1128.7 & 128 & 2 & 0.83(5) & 27.2(8)  \\
W & pI & 1064.1 & 128 & 2 &  0.73(5) & 29.5(8) \\
W & pII & 921.55 & 128 & 2 & 0.94(5) & 30.0(8) \\
W & $\gamma_0$ &  &  &  & & 28.9(8)    \\
\hline
\hline
\end{tabular}
\caption{Computational details of the Mold Integration calculations performed to obtain the ice-solution
interfacial free energy for different crystal orientations. For comparison, we also
report the results for pure water (W) from Ref. \cite{pocillosagua}. $\gamma_0$ is the 
average of the three orientations. $N_l$ is the number of crystal layers in the mold.}
\label{tablapocillos}
\end{table}

\begin{figure}[!htb]
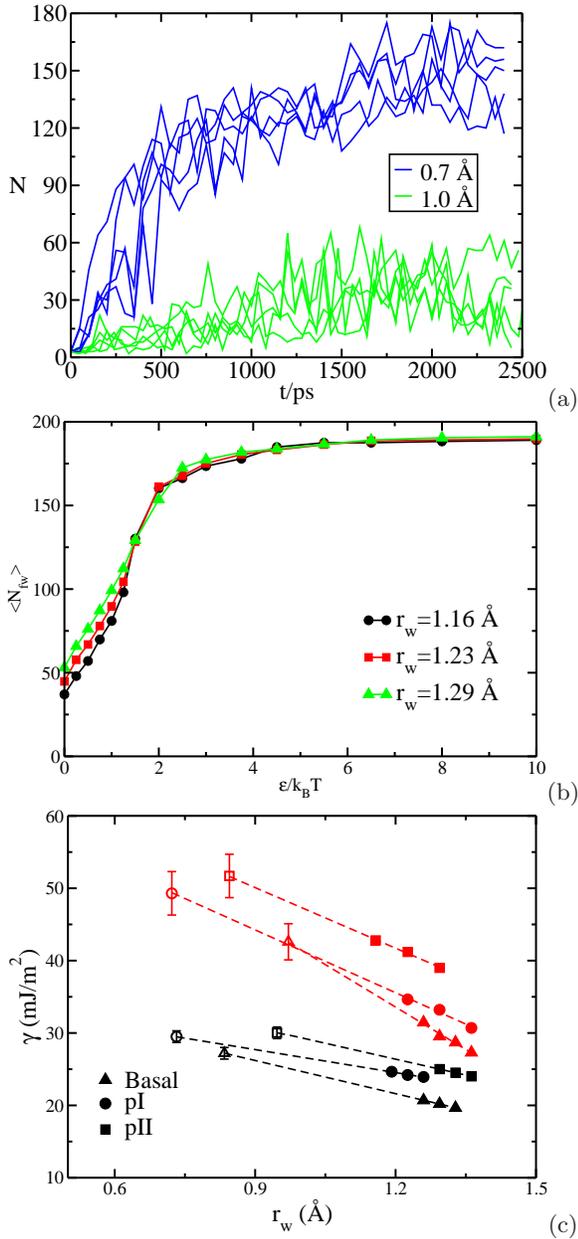

\begin{center}
\includegraphics[clip,width=0.4\textwidth]{./width_choice_secondary_prism_face.eps}(a)
\includegraphics[clip,width=0.4\textwidth]{./integrales_priismatic_secondary.eps}(b)
\includegraphics[clip,width=0.4\textwidth]{./gammas_tip4p2005.eps}(c)
\caption{(a) Number of ice-like molecules growing from the mold as a function of time for several trajectories starting from the studied NaCl
solution at coexistence conditions. The mold is switched on at the beginning of the simulation. The well radious $r_w$
is 1 \AA~ for the green curves and 0.7 \AA~ for the blue ones. (b) Integrand of Eq. \ref{eqpoc} for several $r_w$ values, as indicated in 
the legend, for the secondary prismatic plane. (c) Ice-fluid interfacial free energy as a function of $r_w$ for several
crystal orientations as indicated in the legend. Red data correspond to
the solution and black ones, from Ref. \cite{pocillosagua}, to pure water. Filled symbols are the results of our calculations and 
empty ones are the extrapolation to the corresponding $r_w^o$, which give the final $\gamma$ values.}
\label{figpoc}
\end{center}
\end{figure}

In order to find $r_w^o$ we run several simulations for different well radii
monitoring the number of ice-like molecules.  
The simulations start from a configuration of the solution at coexistence conditions
(1 bar and 240.5 K) and the mold is switched on at the beginning of the simulation.
$r_w^o$ is enclosed between the
largest radious for which ice grows with no induction period and the smallest 
one for which ice either does not grow, or grows after some induction time
\cite{espinosaJCP2014_2}. Such runs for the pII plane are shown in Fig.
\ref{figpoc}(a). For $r_w=0.7$\AA~ (blue curves) or smaller (not shown) ice grows
with no induction period in all trajectories, whereas for $r_w=1$\AA~ (green
curves) or larger (not shown) ice does not grow.  Thus, for this case we set
$r_w^o=0.85\pm0.10$\AA. 

Once we set a value for $r_w^o$ we obtain
$\Delta G$ from Eq. \ref{eqpoc} for several values of $r_w$ larger
than $r_w^o$. The integrand of Eq. \ref{eqpoc} for the pII orientation is shown in Fig. \ref{figpoc}(b) for
several $r_w$ values. Each point in Fig. \ref{figpoc}(b) is obtained in a simulation with well depth $\epsilon$
(indicated by the x-axis in the figure). 
By integrating these curves we obtain $\gamma(r_w)=\Delta G(r_w)/(2A)$, as shown in Fig. \ref{figpoc}(c)
for all studied orientations. Filled symbols correspond to our calculations for $r_w>r_w^o$ and 
empty ones to a linear extrapolation to $r_w=r_w^o$, that gives the value for $\gamma$. 
As in the case of pure water, the interfacial free energy of the prismatic planes is higher than 
that of the basal plane. In fact, the $\gamma$ anisotropy is higher for salty than for pure water: 
in the case of pure water the interfacial free energy of the prismatic planes is about
2-3 mN/m higher than that of the basal plane, whereas for salty water the difference is about 6-9 mN/m. 
For all studied planes, the solid-fluid interfacial free energy is higher 
in the solution than in pure water. 
Therefore, the Mold Integration method confirms the seeding prediction that the
ice-fluid interfacial free energy is larger for the solution. 
The orientationally averaged $\gamma$ from the Mold Integration
calculations is plotted alongside the seeding results in Fig.
\ref{various}(c). Within our statistical uncertainty, the
value from Mold Integration ($\Delta T=0$) is consistent with the seeding calculations ($\Delta T>0$).

\subsection{Nucleation rate}

\begin{figure}[!htb]
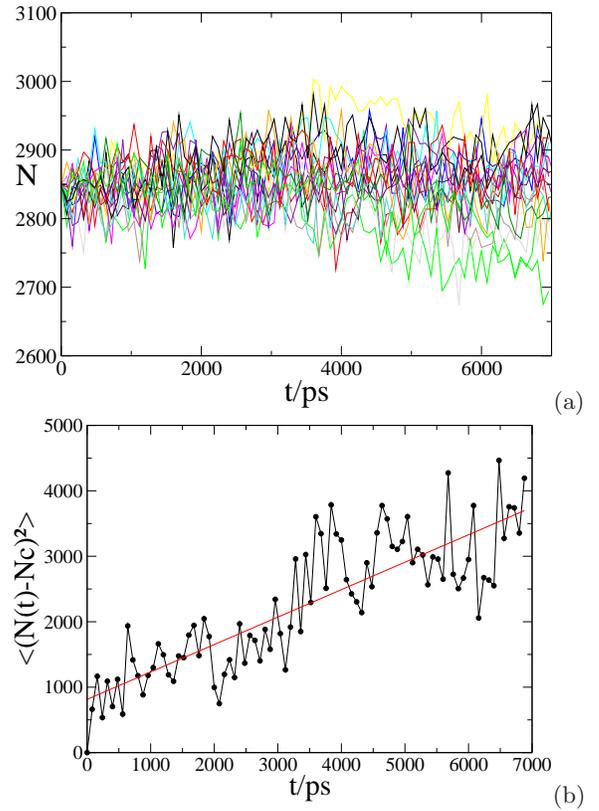

\begin{center}
\includegraphics[clip,width=0.4\textwidth]{./18_semillas.eps}(a)
\includegraphics[clip,width=0.4\textwidth]{./Att_rate_18semillas.eps}(b)
\caption{(a) Time evolution of the number of particles, $N$, in a critical ice cluster of $\sim$ 2850 molecules in 
salty water. Different trajectories correspond to runs launched with different velocity distributions. (b) 
Mean squared difference of the number of molecules in the cluster at time t and time 0 as a function of time.} 
\label{att}
\end{center}
\end{figure}

\begin{figure}[!htb]
\begin{center}
\includegraphics[clip,width=0.4\textwidth]{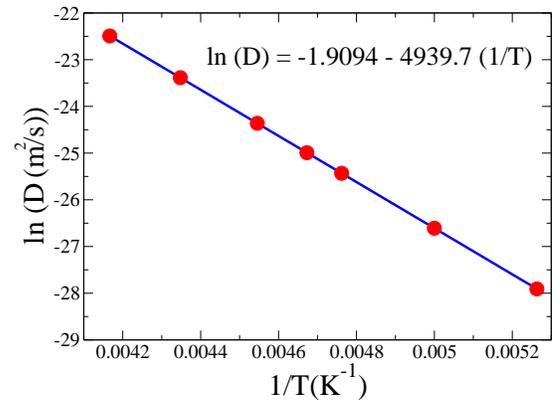}
\caption{Diffussion coefficient of water in the salty solution as a function of the
inverse temperature.}
\label{D}
\end{center}
\end{figure}

We have already shown in Fig. \ref{various}(a) that, for a given $\Delta T$, the size of 
critical ice clusters is larger in salty than in pure water. In the previous section we
argue that this is due to the increase of the ice-liquid interfacial free energy. In the
present section we aim at quantifying the extent to which ice nucleation is slowed down
by adding salt. The speed of ice nucleation is measured by the nucleation rate, $J$,
that is simply defined as the number of critical ice clusters appearing per unit time and volume. 
According to CNT $J$ is given by:
\begin{equation}
J=A e^{-\Delta G_c/(k_BT)}, 
\label{rateeq}
\end{equation}
where $\Delta G_c$ has already been computed from Eq. \ref{dgceq}  
(see Fig. \ref{various}(d)). $A$, the kinetic pre-factor, is given by:
\begin{equation}
A=\rho_w\sqrt{\frac{|\Delta \mu_w|}{6\pi k_BT N_c}}f^+
\label{kinprefeq}
\end{equation}
where $\rho_w$ is the density of water in the solution and $f^+$ is the attachment rate, 
or the frequency with which new water molecules attach to the critical cluster, that can 
be calculated as \cite{Nature_2001_409_1020} 
\begin{equation}
f^+=<(N(t)-N_c)^2/(2t)>
\label{atteq}
\end{equation}
Computing $f^+$ requires performing several simulations of the critical cluster where
$N$ is monitored as a function of time, as shown in Fig. \ref{att}(a).  By averaging $(N(t)-N_c)^2$ over all these trajectories
we obtain a curve such as that shown in Fig. \ref{att}(b), whose slope is 2$f^+$. 

The calculation of $f^+$ is rather involving and we have only performed it as described above for the
cluster containing $N_c=2850$ molecules. For the other two studied clusters we have used the following expression
for $f^+$ provided by CNT:
\begin{equation}
f^+=\frac{24DN_c^{2/3}}{\lambda^2}
\label{atteqap}
\end{equation}
where $D$ is the diffusion coefficient of water in solution, which we plot as a function of 
temperature in Fig. \ref{D}, and $\lambda$ is the 
distance travelled by particles in the vicinity of the cluster's surface to attach
to the cluster. By equating the value of $f^+$ obtained via Eq. \ref{atteq}
for the cluster with 2850 molecules to Eq. \ref{atteqap}
we obtain $\lambda=6$\AA, which is a reasonable value of the order of the molecular diameter. 
We use this value of $\lambda$ combined with Eq. \ref{atteqap} to estimate $f^+$ 
for the other two studied clusters. The values of $f^+$ thus obtained are reported in Table \ref{tabla}. 

We now have all factors required to compute the nucleation rate via Eq. \ref{rateeq}.
The results for the three studied clusters are reported in table \ref{tabla} and 
plotted in Fig. \ref{various}(e) in comparison to those of pure water from Refs. \cite{jacs2013,espinosaJCP2014}.
For a given supercooling, the nucleation rate is lower in salty than in pure water,
as expected from the result that the critical cluster size is larger in solution (Fig. \ref{various}(a)).

\subsection{Decrease of $J$ when adding salt}

\begin{figure}[!htb]
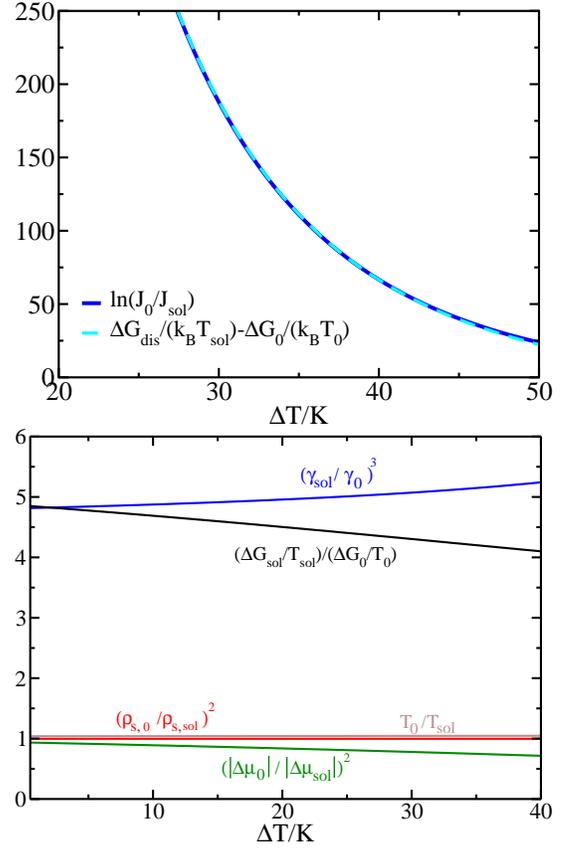

\begin{center}
\includegraphics[clip,width=0.4\textwidth]{./figura_10a.eps}
\includegraphics[clip,width=0.4\textwidth]{./cocientes.eps}
\caption{(a) Difference in $\ln J$ and $-\Delta G_c/(k_BT)$ between salty and pure water as a function of the supercooling.
(b) Black curve, factor by which $\Delta G_c/(k_BT)$ increases in the salty solution with respect to pure water as a function of the supercooling.
Coloured curves, different factors that contribute to such increase, as indicated
in the figure. The product of the coloured curves gives
the black curve.} 
\label{factors}
\end{center}
\end{figure}

We now try to better rationalise which are the factors that contribute to the
decrease of the nucleation rate when adding salt.  To do that we first need to
fit the data coming from the seeding simulations, which we do using the CNT
expressions above \cite{seedingvienes,espinosaJCP2014}. 
In our fitting procedure we assume a
linear temperature dependence of $\gamma$.
We obtain the linear fit by combining seeding and Mold Integration data
(see Fig. \ref{various}(c)).
With $\gamma(T)$ and Eq. \ref{gammaeq} we obtain $N(T)$ 
(solid lines in Fig. \ref{various}(a)). From $N(T)$, $\Delta \mu_w (T)$ and
Eq. \ref{dgceq} we obtain CNT fits to $\Delta G_c(T)$ (solid lines in Fig. \ref{various}(d)).
Finally, with the Arrhenius-like fit to $D(T)$, shown in
Fig. \ref{D}, and Eqs. \ref{atteqap} and \ref{rateeq} we get the fits to $J$ shown  
in Fig. \ref{various}(e).

Having the temperature dependence of the factors that affect the nucleation
rate we can quantify  and compare their variation when adding salt. 
We perform the comparison at constant supercooling rather than at absolute temperature
to get rid of the trivial lowering in $J$ caused by the decrease of the
melting temperature. 
We compare
first the extent to which the kinetic prefactor
and $\Delta G_c/(k_BT)$ affect the drop in $J$. 
According to Eq.
\ref{rateeq} the difference in
$\ln (J)$ between pure (0) and salty (sol)  water,
$\ln(J_{0}/J_{sol})$, has two terms, one for the kinetic
pre-factor, $\ln (A_{0}/A_{sol})$, and one for the free energy barrier
height, $\Delta G^{sol}_c(T_{sol})/(k_BT_{sol})-\Delta G^{0}_c(T_{0})/(k_BT_{0})$ (note that $T_0$ and $T_{sol}$ are not
the same because we compare at constant $\Delta T$).  The fact that
$\ln(J_{0}/J_{sol})$ is equal to the difference in $\Delta G_c/(k_BT)$ (see Fig. \ref{factors}(a))
means that the kinetic prefactor is not significantly affected by salt. 
It is then ratio between the free energy barrier height and the thermal energy that changes when adding salt. 
In Fig. \ref{factors}(b), black curve, we show that $\Delta G_c/(k_BT)$ is between 4 and 5 times
larger in the salty solution. 
According to Eq. \ref{dgceq} the factors that affect $\Delta G_c/(k_BT)$ are 
$\Delta \mu_{w}$, $T$, $\rho_s$ and $\gamma$. In Fig. \ref{factors}(b) we 
show the factor by which each of them contribute to the change of $\Delta G_c/(k_BT)$ when 
adding salt. The ice density and the temperature have a negligible effect
on the change of the free energy barrier height (red and brown curves in Fig. \ref{factors}(b)). The chemical potential difference 
does not change the barrier at low supercooling and lowers it about thirty per cent
at high supercooling (green curve in Fig. \ref{factors}(b)). As previously discussed, 
such lowering is a consequence of the fact that the  
partial molar enthalpy of water in solution is higher than in pure water. Therefore, 
$\Delta \mu_{w}$ aids ice nucleation when salt is added. However, 
the interfacial free energy largely compensates the modest effect of 
$\Delta \mu_{w}$, given that the increase of $\gamma$ 
with salt
multiplies by a factor of 5 the free energy barrier (blue curve in Fig. \ref{factors}(b)). The effect of the 
increase of $\gamma$  on $\Delta G_c/(k_BT)$ is magnified by the fact that $\gamma$ goes as a third power
in Eq. \ref{dgceq}. 
In summary, the driving force for ice nucleation, $\Delta \mu_{w}$, increases when adding salt, however
$\gamma$, that hinders ice nucleation, increases to a greater extent, which causes 
the salt-induced deceleration of ice nucleation.

\section{Discussion}

\subsection{Comparison with the experiment}

\subsubsection{Melting point depression}

Experimentally, a 1.85 m NaCl solution freezes 6.6 K below the melting point of
pure water \cite{descensocrios}.  The model captures that salt decreases the ice melting
temperature, predicting a 9.5 K depression for a 1.85 m solution. Such depression is caused in experiments by a 
2.6 m NaCl concentration \cite{descensocrios}.  Hence, the effect of salt on the melting point is
larger in the model than in the experiment. 
This suggests that the NaCl model we are using 
causes a larger decrease in the chemical potential of TIP4P/2005 water 
than that caused by real NaCl in real water (see Fig. 4b of Ref. \cite{analaura}). 

To account for such enhanced effect of the model, it is interesting to compare model and experiment
for the same activity of water in the solution coexisting with ice, $a_w^i$. 
We can compute $a_w^i$ as:
\begin{equation}
a_w^i=\exp(\Delta \mu_0/(RT))
\label{awie}
\end{equation}
where $\Delta \mu_0$ is the water chemical potential difference
in ice and in pure liquid water at the ice-solution coexistence temperature ($T_m=$240.5K). 
We know $\Delta \mu_0$ from our previous work on homogeneous ice nucleation \cite{jacs2013,espinosaJCP2014}.
We obtain $a_w^i=$0.92. For such $a_w^i$, a melting point of $\sim$ 263 K can be interpolated
from the experimental data reported in Ref. \cite{alpertNaClaqueous}, which is
about 10 K below the melting temperature of pure water. Such melting point depression of 
10 K compares better with the 9.5 K obtained with our model. 
Therefore, when experiment and model are compared for a given water activity, a better
agreement is obtained than when they are compared at a given concentration.
In order to improve the employed model for NaCl aqueous solutions 
it is therefore necessary to modify the ion-water interactions in such way that a smaller
water activity drop is caused by the ions.

\subsubsection{Nucleation rate}

It is experimentally known that the supercooling required to freeze microscopic salty water
drops is larger than that required to freeze pure water drops \cite{kannoJPC1977,koopNature2000,alpertNaClaqueous}.
Equivalently, for a given supercooling, the nucleation rate is higher in pure than in salty water. 
In  Fig. \ref{various}(e) we show that the model indeed captures the experimental trend. 
According to our results, such trend is due to an increase of the ice-solution
interfacial free energy when adding salt. This simulation prediction is
quite valuable considering the  
lack of accuracy in experimental measurements/estimates of the 
ice-liquid interfacial free energy  \cite{ickespccp2015}. 

According to the employed model (Fig. \ref{various}(e)), the nucleation rates in pure and salty water are quite similar at very high supercooling (about 60 K). 
The rate in such conditions can be measured using nanoscopic drops \cite{manka2012}. 
Consistently with the similarity of the rate, the interfacial free energy of pure water is also predicted to be similar to that of salty water at high 
supercooling (see Fig. \ref{various}(c)).  

The model predictions for the nucleation rate in salty water, shown in Fig. \ref{various}(e),
can be directly compared to experimental measurements by Alpert et al. \cite{alpertNaClaqueous} (empty symbols). 
The comparison can be made either for a solution with the same NaCl concentration (empty circle)
or for one with the same water activity at the melting temperature, $a_w(T_m)$ (empty diamond).
In either case  
the agreement  between simulation and experiment is quite
satisfactory.


\section{Summary and Conclusions}

We use computer simulations to investigate the effect of salt on homogeneous
ice nucleation.  To study the aqueous solution, we use the TIP4P/2005 model for water in combination with the
Joung Cheetham NaCl model as parametrised for SPC/E water. To start with, we compute the ice
melting point for the model using both direct coexistence and thermodynamic
integration from previous calculations of the chemical potential of water in
solution.  The model predicts a larger cryoscopic depression than 
the experimental one for a given
salt concentration. 
However, if we compare experiment and simulation for the same
activity of water at coexistence we obtain a good agreement for the melting
point drop.  This suggests that
the employed NaCl model is affecting the solute (water) to a greater extent than
real NaCl does.   

After computing the melting temperature we compute the size of
critical ice clusters by embedding spherical ice Ih seeds in the supercooled
solution and simulating trajectories at different temperatures. We compute the
chemical potential difference of water in ice and in the solution using
thermodynamic integration of the partial molar enthalpy from the melting
temperature. With such chemical potential difference and the number of
particles in the critical cluster we obtain the height of the ice nucleation
free energy barrier and the ice-solution interfacial free energy using the
expressions provided by Classical Nucleation Theory.  
We also compute the
ice-solution interfacial free energy at coexistence using the Mold Integration
method, that provides a thermodynamic route to reversibly grow an ice slab in
the fluid at coexistence conditions.   The interfacial free energies obtained
with seeding in supercooled conditions extrapolate linearly to the value at
coexistence obtained with Mold Integration.  This consistency test supports the
validity of our approach.
The interfacial free energy at
any supercooling increases with respect to that of pure
water (obtained in previous studies). 

The kinetic pre-factor for the ice nucleation rate
is obtained for one of the ice seeds by launching many trajectories and
computing the mean squared displacement of the number of molecules in the
cluster. 
The Classical Nucleation Theory expression for the kinetic pre-factor, which is
proportional to the diffusion coefficient of water molecules in the liquid, 
is consistent with such calculation. 
Such consistency enabled us  to use the diffusion coefficient of water as a function of temperature
to get the temperature dependence of the kinetic pre-factor. 
That, combined with the temperature
dependence obtained for the ice-solution chemical potential difference of water
and for the interfacial free energy, gave us the temperature dependence of the
nucleation rate for a wide range of orders of magnitude.  

The model
qualitatively captures the experimental trend that, for a given supercooling,
the nucleation rate decreases by adding salt. Our model predicts that salt
hinders ice nucleation despite the fact that the chemical potential difference
between water in ice and in solution is larger when salt is added.  This would
in principle favour ice nucleation for a certain supercooling. However, the increase of the interfacial
free energy largely compensates for the increase of the thermodynamic driving
force for nucleation and the net effect is a deceleration of the ice nucleation
process. The ice nucleation rate predicted by the salty water model  is in good
agreement with experimental measurements, which brings confidence in the
predictions made by the model.

\textbf{Acknowledgements}

This work was funded by grants FIS2013/43209-P and FIS2016-78117-P FIS2016-78847-P of the MEC, and the
UCM/Santander 910570 and PR26/16-10B-2. C.Valeriani and E. Sanz acknowledge financial support
from a Ramon y Cajal Fellowship. J.  R. Espinosa acknowledges financial support
from the FPI grant BES-2014-067625.  Calculations  were carried out in the
supercomputer facilities  La Palma and Magerit  from the Spanish Supercomputing Network (RES)
(projects QCM-2015-3-0036 and QCM-2016-1-0039).
The authors acknowledge the computer resources and technical assistance
provided by the Centro de Supercomputacion y Visualizacion de Madrid (CeSViMa).

\clearpage

\appendix
\section{Determining $N$}
\label{orderparam}

\begin{figure}[!htb]
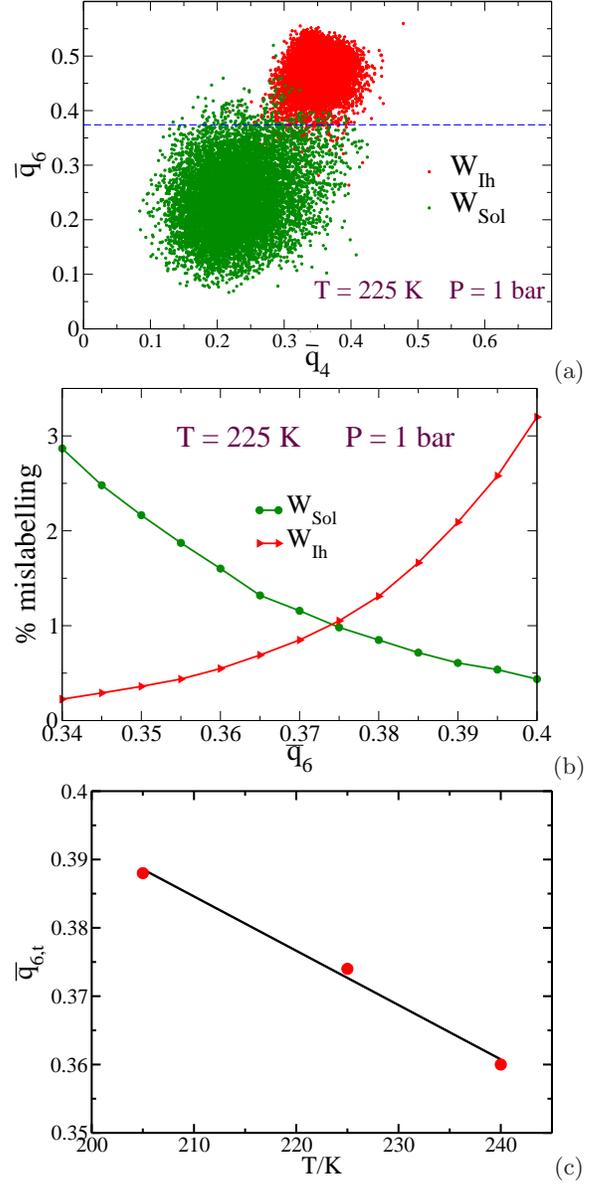

\begin{center}
\includegraphics[clip,width=0.4\textwidth]{./nubes.eps}(a)
\includegraphics[clip,width=0.4\textwidth]{./curva_mislabelling.eps}(b)
\includegraphics[clip,width=0.4\textwidth]{./umbral.eps}(c)
\caption{(a) $\bar{q}_{6}$ vs $\bar{q}_{4}$ for water in the salty solution (green) and water in ice Ih (red). (b) Fraction of mislabelled particles as a 
	function of the $\bar{q}_{6}$ threshold for both phases at 225 K and 1 bar. The $\bar{q}_{6}$ threshold is chosen at the crossing point between
both curves. (c) Selected $\bar{q}_{6}$ threshold, $\bar{q}_{6,t}$, as a function of temperature.}
\label{po}
\end{center}
\end{figure}

To determine the number of particles in the ice phase (be it the ice slab
in the direct coexistence simulations or the cluster in the seeding ones) 
we use the $\bar{q}_{i}$
rotationally invariant local-bond order parameter proposed in Ref.
\cite{lechnerJCP2008}. $\bar{q}_{i}$ is a scalar number
whose value for a given particle
depends on the relative positions of the tagged particle and its neighbors within a
certain distance (3.5 \AA, around the first minimum in the oxygen-oxygen radial distribution
function of the liquid phase). We only use the oxygens for the calculation of 
$\bar{q}_{i}$. In Fig. \ref{po} (a) we plot typical ($\bar{q}_{4}$,$\bar{q}_{6}$) 
values for ice (red) and solution (green) water molecules. Differently from $\bar{q}_{4}$,
it is possible to establish a threshold for $\bar{q}_{6}$, $\bar{q}_{6,t}$,
to discriminate between ice-like and solution-like water molecules (indicated with a horizontal dashed line in Fig. \ref{po}(a)).
To establish the $\bar{q}_{6,t}$ value we look at the point at which 
the fraction of wrongly labelled particles in both phases coincide (see Fig. \ref{po}(b)). 
As shown in Fig. \ref{po}(c), the value of such threshold depends on temperature. 
For a given temperature, particles with $\bar{q}_{6} > \bar{q}_{6,t}$
are labelled as solid-like. 
Having established all solid-like particles in the system, we
cluster them using a neighbour cut-off distance of 3.5\AA.
$N$ is the number of molecules in the largest detected cluster of solid-like molecules.

\clearpage


\begin{thebibliography}{10}

\bibitem{weathering_book}
A.~G. Gerrard, {\em Rocks and Landforms}.
\newblock Springer Netherlands, 1988.

\bibitem{bacterial_ice_nucleation}
J.~K. Li and T.~C. Lee, ``Bacterial ice nucleation and its potential
  application in the food-industry,'' {\em Trends in Food Sci. \& Tech.},
  vol.~6, pp.~259--265, 1995.

\bibitem{maki74}
L.~R. Maki, E.~L. Galyan, C.~M.M., and D.~R. Caldwell, ``Ice nucleation induced
  by pseudomonas-syringae,'' {\em Appl. Microbiol.}, vol.~28, pp.~456--459,
  1974.

\bibitem{cryopres}
G.~J. Morris and E.~Acton, ``Controlled ice nucleation in cryopreservation--a
  review,'' {\em Cryobiology}, vol.~66, p.~85, 2013.

\bibitem{kannoJPC1977}
H.~Kanno and C.~A. Angell, ``Homogeneous nucleation and glass formation in
  aqueous alkali halide solutions at high pressures,'' {\em The Journal of
  Physical Chemistry}, vol.~81, no.~26, pp.~2639--2643, 1977.

\bibitem{kelton}
K.~F. Kelton, {\em Crystal Nucleation in Liquids and Glasses}.
\newblock Boston: Academic, 1991.

\bibitem{alpertNaClaqueous}
P.~A. Alpert, J.~Y. Aller, and D.~A. Knopf, ``Ice nucleation from aqueous nacl
  droplets with and without marine diatoms,'' {\em Atmos. Chem. Phys.},
  vol.~11, p.~5539, 2011.

\bibitem{ickespccp2015}
L.~Ickes, A.~Welti, C.~Hoose, and U.~Lohmann, ``Classical nucleation theory of
  homogeneous freezing of water: thermodynamic and kinetic parameters,'' {\em
  Phys. Chem. Chem. Phys.}, vol.~17, pp.~5514--5537, 2015.

\bibitem{reviewMichaelides2016}
G.~C. Sosso, J.~Chen, S.~J. Cox, M.~Fitzner, P.~Pedevilla, A.~Zen, and
  A.~Michaelides, ``Crystal nucleation in liquids: Open questions and future
  challenges in molecular dynamics simulations,'' {\em Chemical Reviews},
  vol.~116, no.~12, pp.~7078--7116, 2016.

\bibitem{JACS_2003_125_07743}
R.~Radhakrishnan and B.~L. Trout, ``Nucleation of hexagonal ice ({I}h) in
  liquid water,'' {\em J. Am. Chem. Soc.}, vol.~125, p.~7743, 2003.

\bibitem{quigley:154518}
D.~Quigley and P.~M. Rodger, ``Metadynamics simulations of ice nucleation and
  growth,'' {\em J. Chem. Phys.}, vol.~128, no.~15, p.~154518, 2008.

\bibitem{geigerJCP2013}
P.~Geiger and C.~Dellago, ``Neural networks for local structure detection in
  polymorphic systems,'' {\em J. Chem. Phys.}, vol.~139, no.~16, p.~164105,
  2013.

\bibitem{valeria_pccp_2011}
E.~B. Moore and V.~Molinero, ``Is it cubic? ice crystallization from deeply
  supercooled water,'' {\em Phys. Chem. Chem. Phys.}, vol.~13, p.~20008, 2011.

\bibitem{Malkinpnas2012}
T.~L. Malkin, B.~J. Murray, A.~V. Brukhno, J.~Anwar, and C.~G. Salzmann,
  ``Structure of ice crystallized from supercooled water,'' {\em Proceedings of
  the National Academy of Sciences}, vol.~109, no.~4, pp.~1041--1045, 2012.

\bibitem{galli_mw}
T.~Li, D.~Donadio, G.~Russo, and G.~Galli, ``Homogeneous ice nucleation from
  supercooled water,'' {\em Phys. Chem. Chem. Phys.}, vol.~13,
  pp.~19807--19813, 2011.

\bibitem{haji-akbariPNAS2015}
A.~Haji-Akbari and P.~G. Debenedetti, ``Direct calculation of ice homogeneous
  nucleation rate for a molecular model of water,'' {\em Proceedings of the
  National Academy of Sciences}, vol.~112, no.~34, pp.~10582--10588, 2015.

\bibitem{jacs2013}
E.~Sanz, C.~Vega, J.~R. Espinosa, R.~Caballero-Bernal, J.~L.~F. Abascal, and
  C.~Valeriani, ``Homogeneous ice nucleation at moderate supercooling from
  molecular simulation,'' {\em Journal of the American Chemical Society},
  vol.~135, no.~40, pp.~15008--15017, 2013.

\bibitem{chakrabortyJPCL2013}
D.~Chakraborty and G.~N. Patey, ``How crystals nucleate and grow in aqueous
  nacl solution,'' {\em The Journal of Physical Chemistry Letters}, vol.~4,
  no.~4, pp.~573--578, 2013.

\bibitem{lanaroNaCl2016}
G.~Lanaro and G.~N. Patey, ``Birth of nacl crystals: Insights from molecular
  simulations,'' {\em The Journal of Physical Chemistry B}, vol.~120, no.~34,
  pp.~9076--9087, 2016.

\bibitem{PRL_2004_92_040801}
D.~Zahn, ``Atomistic mechanism of {N}a{C}l nucleation from an aqueous
  solution,'' {\em Phys. Rev. Lett.}, vol.~92, p.~040801, 2004.

\bibitem{petersJACS2015}
N.~E.~R. Zimmermann, B.~Vorselaars, D.~Quigley, and B.~Peters, ``Nucleation of
  {N}a{C}l from aqueous solution: Critical sizes, ion-attachment kinetics, and
  rates,'' {\em Journal of the American Chemical Society}, vol.~137, no.~41,
  pp.~13352--13361, 2015.

\bibitem{condePCCP2017}
M.~M. Conde, M.~Rovere, and P.~Gallo, ``Spontaneous nacl-doped ice at seawater
  conditions: focus on the mechanisms of ion inclusion,'' {\em Phys. Chem.
  Chem. Phys.}, vol.~19, pp.~9566--9574, 2017.

\bibitem{bullockFD2013}
G.~Bullock and V.~Molinero, ``Low-density liquid water is the mother of ice: on
  the relation between mesostructure{,} thermodynamics and ice crystallization
  in solutions,'' {\em Faraday Discuss.}, vol.~167, pp.~371--388, 2013.

\bibitem{icenucNaClJung2008}
S.~Bauerecker, P.~Ulbig, V.~Buch, L.~Vrbka, and P.~Jungwirth, ``Monitoring ice
  nucleation in pure and salty water via high-speed imaging and computer
  simulations,'' {\em The Journal of Physical Chemistry C}, vol.~112, no.~20,
  pp.~7631--7636, 2008.

\bibitem{kannoScience1975}
H.~Kanno, R.~J. Speedy, and C.~A. Angell, ``Supercooling of water to -92$^o$c
  under pressure,'' {\em Science}, vol.~189, no.~4206, pp.~880--881, 1975.

\bibitem{espinosaPRL2016}
J.~R. Espinosa, A.~Zaragoza, P.~Rosales-Pelaez, C.~Navarro, C.~Valeriani,
  C.~Vega, and E.~Sanz, ``Interfacial free energy as the key to the
  pressure-induced deceleration of ice nucleation,'' {\em Phys. Rev. Lett.},
  vol.~117, p.~135702, 2016.

\bibitem{espinosaJPCL2017}
J.~R. Espinosa, G.~D. Soria, J.~Ramirez, C.~Valeriani, C.~Vega, and E.~Sanz,
  ``Role of salt, pressure, and water activity on homogeneous ice nucleation,''
  {\em J. Phys. Chem. Lett.}, vol.~8, p.~4486, 2017.

\bibitem{JCP_2005_123_234505}
J.~L.~F. Abascal and C.~Vega, ``A general purpose model for the condensed
  phases of water: {TIP4P}/2005,'' {\em J. Chem. Phys.}, vol.~123, p.~234505,
  2005.

\bibitem{JCNaClpotential}
I.~S. Joung and T.~E. Cheatham, ``Determination of alkali and halide monovalent
  ion parameters for use in explicitly solvated biomolecular simulations,''
  {\em The Journal of Physical Chemistry B}, vol.~112, no.~30, pp.~9020--9041,
  2008.

\bibitem{analaura}
A.~L. Benavides, J.~L. Aragones, and C.~Vega, ``Consensus on the solubility of
  {N}a{C}l in water from computer simulations using the chemical potential
  route,'' {\em The Journal of Chemical Physics}, vol.~144, p.~124504, 2016.

\bibitem{ZPC_1926_119_277_nolotengo}
M.~Volmer and A.~Weber, ``Keimbildung in ubersattigten gebilden,'' {\em Z.
  Phys. Chem.}, vol.~119, p.~277, 1926.

\bibitem{becker-doring}
R.~Becker and W.~Doring, ``Kinetische behandlung der keimbildung in
  ubersattigten dampfen,'' {\em Ann. Phys.}, vol.~416, pp.~719--752, 1935.

\bibitem{laddcpl1977}
A.~Ladd and L.~Woodcock, ``Triple-point coexistence properties of the
  lennard-jones system,'' {\em Chemical Physics Letters}, vol.~51, no.~1,
  pp.~155 -- 159, 1977.

\bibitem{JCP_2007_126_014507}
E.~Sanz and C.~Vega, ``Solubility of {KF} and {NaCl} in water by molecular
  simulation,'' {\em J. Chem. Phys.}, vol.~126, p.~014507, 2007.

\bibitem{frenkelbook}
D.~Frenkel and B.~Smit, {\em Understanding Molecular Simulation}.
\newblock Academic Press, London, 1996.

\bibitem{seedingvienes}
J.~R. Espinosa, C.~Vega, C.~Valeriani, and E.~Sanz, ``Seeding approach to
  crystal nucleation,'' {\em J. Chem. Phys.}, vol.~144, p.~034501, 2016.

\bibitem{Nature_2001_409_1020}
S.~Auer and D.~Frenkel, ``Prediction of absolute crystal-nucleation rate in
  hard-sphere colloids,'' {\em Nature}, vol.~409, p.~1020, 2001.

\bibitem{espinosaJCP2014_2}
J.~R. Espinosa, C.~Vega, and E.~Sanz, ``The mold integration method for the
  calculation of the crystal-fluid interfacial free energy from simulations,''
  {\em J Chem. Phys.}, vol.~141, no.~13, p.~134709, 2014.

\bibitem{pocillosagua}
J.~R. Espinosa, C.~Vega, and E.~Sanz, ``Ice-water interfacial free energy for
  the tip4p, tip4p/2005, tip4p/ice and mw models as obtained from the mold
  integration technique,'' {\em The Journal of Physical Chemistry C}, vol.~120,
  pp.~8068--8075, 2016.

\bibitem{espinosaJCP2014}
J.~R. Espinosa, E.~Sanz, C.~Valeriani, and C.~Vega, ``Homogeneous ice
  nucleation evaluated for several water models,'' {\em J. Chem. Phys.},
  vol.~141, p.~18C529, 2014.

\bibitem{lorentzrule}
H.~A. Lorentz, ``Ueber die anwendung des satzes vom virial in der kinetischen
  theorie der gase,'' {\em Annalen der Physik}, vol.~248, no.~1, pp.~127--136,
  1881.

\bibitem{berthelotrule}
D.~Berthelot {\em C. R. Acad. Sci.}, vol.~126, p.~1713, 1898.

\bibitem{solubNaCl}
W.~J. Hamer and Y.-C. Wu, ``Osmotic coefficients and mean activity coefficients
  of uni¿univalent electrolytes in water at 25°c,'' {\em Journal of Physical
  and Chemical Reference Data}, vol.~1, no.~4, pp.~1047--1100, 1972.

\bibitem{hess08}
B.~Hess, C.~Kutzner, D.~van~der Spoel, and E.~Lindahl, ``Algorithms for highly
  efficient, load-balanced, and scalable molecular simulation,'' {\em J. Chem.
  Theory Comput.}, vol.~4, pp.~435--447, 2008.

\bibitem{parrinello81}
M.~Parrinello and A.~Rahman, ``Polymorphic transitions in single crystals: A
  new {M}olecular {D}ynamics method,'' {\em J. App. Phys.}, vol.~52,
  pp.~7182--7190, 1981.

\bibitem{bussi07}
G.~Bussi, D.~Donadio, and M.~Parrinello, ``Canonical sampling through velocity
  rescaling,'' {\em The Journal of Chemical Physics}, vol.~126, no.~1,
  p.~014101, 2007.

\bibitem{CPL_2002_366_0537}
D.~R. Wheeler and J.~Newman, ``A less expensive ewald lattice sum,'' {\em Chem.
  Phys. Lett.}, vol.~366, p.~537, 2002.

\bibitem{lincs97}
B.~Hess, H.~Bekker, H.~J.~C. Berendsen, and J.~G. E.~M. Fraaije, ``Lincs: A
  linear constraint solver for molecular simulations,'' {\em J. Comput. Chem.},
  vol.~18, no.~12, pp.~1463--1472, 1997.

\bibitem{lincs2008}
B.~Hess, ``P-lincs:¿ a parallel linear constraint solver for molecular
  simulation,'' {\em Journal of Chemical Theory and Computation}, vol.~4,
  no.~1, pp.~116--122, 2008.

\bibitem{garciafernandezJCP2005}
R.~Garcia-Fernandez, J.~L.~F. Abascal, and C.~Vega, ``The melting point of ice
  ih for common water models calculated from direct coexistence of the
  solid-liquid interface,'' {\em The Journal of Chemical Physics}, vol.~124,
  no.~14, p.~144506, 2006.

\bibitem{noyaJCP2008}
E.~G. Noya, C.~Vega, and E.~de~Miguel, ``Determination of the melting point of
  hard spheres from direct coexistence simulation methods,'' {\em The Journal
  of Chemical Physics}, vol.~128, no.~15, p.~154507, 2008.

\bibitem{sigmoide}
J.~R. Espinosa, E.~Sanz, C.~Valeriani, and C.~Vega, ``On fluid-solid direct
  coexistence simulations: The pseudo-hard sphere model,'' {\em The Journal of
  Chemical Physics}, vol.~139, no.~14, p.~144502, 2013.

\bibitem{aragones12}
J.~L. Aragones, E.~Sanz, C.~Valeriani, and C.~Vega, ``Calculation of the
  melting point of alkali halides by means of computer simulations,'' {\em J.
  Chem. Phys.}, vol.~137, no.~10, p.~104507, 2012.

\bibitem{condeJCP2013}
M.~M. Conde, M.~A. Gonzalez, J.~L.~F. Abascal, and C.~Vega, ``Determining the
  phase diagram of water from direct coexistence simulations: The phase diagram
  of the {TIP4P}/2005 model revisited,'' {\em The Journal of Chemical Physics},
  vol.~139, no.~15, p.~154505, 2013.

\bibitem{mesterJCP2015}
Z.~Mester and A.~Z. Panagiotopoulos, ``Temperature-dependent solubilities and
  mean ionic activity coefficients of alkali halides in water from molecular
  dynamics simulations,'' {\em The Journal of Chemical Physics}, vol.~143,
  no.~4, p.~044505, 2015.

\bibitem{panagiotopoulosjcp2015}
Z.~Mester and A.~Z. Panagiotopoulos, ``Mean ionic activity coefficients in
  aqueous nacl solutions from molecular dynamics simulations,'' {\em The
  Journal of Chemical Physics}, vol.~142, no.~4, p.~044507, 2015.

\bibitem{paluchJCP2010}
A.~S. Paluch, S.~Jayaraman, J.~K. Shah, and E.~J. Maginn, ``A method for
  computing the solubility limit of solids: Application to sodium chloride in
  water and alcohols,'' {\em The Journal of Chemical Physics}, vol.~133,
  no.~12, p.~124504, 2010.

\bibitem{mouckaJPCB2012}
F.~Moucka, M.~Lisal, and W.~R. Smith, ``Molecular simulation of aqueous
  electrolyte solubility. 3. alkali-halide salts and their mixtures in water
  and in hydrochloric acid,'' {\em The Journal of Physical Chemistry B},
  vol.~116, no.~18, pp.~5468--5478, 2012.

\bibitem{ivoreview2016}
I.~Nezbeda, F.~Moucka, and W.~R. Smith, ``Recent progress in molecular
  simulation of aqueous electrolytes: force fields, chemical potentials and
  solubility,'' {\em Molecular Physics}, vol.~114, no.~11, pp.~1665--1690,
  2016.

\bibitem{solubilityfrenkel2017}
L.~Li, T.~Totton, and D.~Frenkel, ``Computational methodology for solubility
  prediction: Application to the sparingly soluble solutes,'' {\em The Journal
  of Chemical Physics}, vol.~146, no.~21, p.~214110, 2017.

\bibitem{carignano}
R.~G. Pereyra, I.~Szleifer, and M.~A. Carignano, ``Temperature dependence of
  ice critical nucleus size,'' {\em J. Chem. Phys.}, vol.~135, p.~034508, 2011.

\bibitem{baiJCP2006}
X.-M. Bai and M.~Li, ``Calculation of solid-liquid interfacial free energy: A
  classical nucleation theory based approach,'' {\em J. Chem. Phys.}, vol.~124,
  no.~12, p.~124707, 2006.

\bibitem{hurtadoseedingising2008}
P.~I. Hurtado, J.~Marro, and P.~L. Garrido, ``Demagnetization via nucleation of
  the nonequilibrium metastable phase in a model of disorder,'' {\em Journal of
  Statistical Physics}, vol.~133, no.~1, pp.~29--58, 2008.

\bibitem{Jacobson_molinero}
V.~Jacobson, L.C.;Molinero, ``Can amorphous nuclei grow crystalline clathrates?
  the size and crystallinity of critical clathrate nuclei,'' {\em
  J.Am.Chem.Soc.}, vol.~133, p.~6458, 2011.

\bibitem{knottJACS2012}
B.~C. Knott, V.~Molinero, M.~F. Doherty, and B.~Peters, ``Homogeneous
  nucleation of methane hydrates: Unrealistic under realistic conditions,''
  {\em J. Am. Chem. Soc.}, vol.~134, pp.~19544--19547, 2012.

\bibitem{espinosaJCP2015}
J.~R. Espinosa, C.~Vega, C.~Valeriani, and E.~Sanz, ``The crystal-fluid
  interfacial free energy and nucleation rate of {N}a{C}l from different
  simulation methods,'' {\em J. Chem. Phys.}, vol.~142, no.~19, p.~194709,
  2015.

\bibitem{zaragozaJCP2015}
A.~Zaragoza, M.~M. Conde, J.~R. Espinosa, C.~Valeriani, C.~Vega, and E.~Sanz,
  ``Competition between ices ih and ic in homogeneous water freezing,'' {\em
  The Journal of Chemical Physics}, vol.~143, no.~13, p.~134504, 2015.

\bibitem{koopNaClphasediag}
T.~Koop, A.~Kapilashrami, L.~T. Molina, and M.~J. Molina, ``Phase transitions
  of sea-salt/water mixtures at low temperatures: Implications for ozone
  chemistry in the polar marine boundary layer,'' {\em Journal of Geophysical
  Research: Atmospheres}, vol.~105, no.~D21, pp.~26393--26402, 2000.

\bibitem{koopNature2000}
T.~Koop, B.~Luo, A.~Tsias, and T.~Peter, ``Water activity as the determinant
  for homogeneous ice nucleation in aqueous solutions,'' {\em Nature},
  vol.~406, pp.~611--614, 2000.

\bibitem{pruppacher1995}
H.~R. Pruppacher, ``A new look at homogeneous ice nucleation in supercooled
  water drops,'' {\em J. Atmosph. Sci.}, vol.~52, p.~1924, 1995.

\bibitem{murray2010}
B.~J. Murray, S.~L. Broadley, T.~W. Wilson, S.~J. Bull, R.~H. Wills, H.~K.
  Christenson, and E.~J. Murray, ``Kinetics of the homogeneous freezing of
  water,'' {\em Phys. Chem. Chem. Phys.}, vol.~12, p.~10380, 2010.

\bibitem{manka2012}
A.~Manka, H.~Pathak, S.~Tanimura, J.~Wolk, R.~Strey, and B.~E. Wyslouzil,
  ``Freezing water in no man's land,'' {\em Phys. Chem. Chem. Phys.}, vol.~14,
  pp.~4505--4516, 2012.

\bibitem{descensocrios}
R.~W. Potter, M.~A. Clynne, and D.~L. Brown, ``Freezing point depression of
  aqueous sodium chloride solutions,'' {\em Scientific Communications},
  vol.~73, no.~2, pp.~284--285, 1978.

\bibitem{lechnerJCP2008}
W.~Lechner and C.~Dellago, ``Accurate determination of crystal structures based
  on averaged local bond order parameters,'' {\em J. Chem. Phys.}, vol.~129,
  no.~11, p.~114707, 2008.

\end{thebibliography}

\end{document}